\documentclass[a4paper,12pt]{article}

\usepackage{graphicx}
\usepackage{amssymb,bm,mathrsfs,bbm,amscd}
\usepackage[tbtags]{amsmath}
\usepackage{epstopdf}

\begin{document}

\begin{center}
{\large\bf Relativistic baryon-baryon interactions in chiral perturbation theory}

{Xiu-Lei Ren$^{1,2}$}, {Kai-Wen Li$^{3}$}, {Li-Sheng Geng$^{3,4,\P}$}

$^1${State Key Laboratory of Nuclear Physics and Technology,
  School of Physics, Peking University, 100871, Beijing, China}

$^2${Institut f\"{u}r Theoretische Physik II, Ruhr-Universit\"{a}t Bochum, D-44780, Bochum, Germany}

$^3${School of Physics and
Nuclear Energy Engineering \& International Research Center for Nuclei and Particles in the Cosmos, Beihang University, 100191, Beijing, China}

$^4${Beijing Key Laboratory of Advanced Nuclear Materials and Physics, Beihang University, 100191, Beijing, China}

$^\P${E-mail: lisheng.geng@buaa.edu.cn}
\end{center}

\centerline{\bf Abstract}
We report on the recent studies of  leading order baryon-baryon interactions in  covariant baryon chiral
perturbation theory. In the strangeness $S=0$ sector, one can achieve a rather good description of the Nijmegen $np$
phase shifts with angular momenta $J\leq 1$, particularly the
$^1S_0$ and $^3P_0$ partial waves,
comparable with the next-to-leading order (NLO) heavy baryon approach.
In the strangeness $S=-1$ hyperon-nucleon sector, the best fit  of the 36 scattering data is similar to the sophisticated phenomenological models and the NLO heavy baryon approach.

\section{Introduction}

In recent years,  it has been shown that  relativistic
effects play an indispensable role in understanding some fine structure of atomic nuclei, such as the nature of
pseudospin symmetry~\cite{Liang:2014dma}.
Being one of the most important theoretical inputs to ({\it ab initio}) few- and many-body calculations,
baryon-baryon interactions have attracted a lot of attention.
Up to now, most baryon-baryon  interactions are
constructed in the non-relativistic (NR) framework, such as the high precision phenomenological nuclear potentials,
or chiral potentials. In the relativistic framework, only two phenomenological formulations (the Bonn potential and the covariant spectator theory) have
been extensively studied.
It is generally accepted that the phenomenological potentials, though extremely useful, cannot be directly connected to the underlying theory of the strong interaction, quantum
chromodynamics (QCD).

Chiral perturbation theory (ChPT)~\cite{Weinberg:1978kz}, as an effective field theory of low-energy QCD, provides a model-independent framework to
study baryon-baryon interactions. After Weinberg's proposal~\cite{Weinberg:1990rz,Weinberg:1991um}, the chiral potentials have been extensively
investigated, especially in the nucleon-nucleon and hyperon-nucleon sectors.
 Most existing chiral potentials are based on the NR heavy baryon
(HB) ChPT~ (see the extensive reviews~\cite{Bedaque:2002mn,Epelbaum:2008ga,Machleidt:2011zz} and references therein). On the other hand, in recent years,
the covariant ChPT has been successfully applied in the one-baryon sector and heavy-light systems,
and has exhibited relatively faster convergence than the HB ChPT.
This encourages us to extend the covariant ChPT to the baryon-baryon sector and to construct  relativistic
baryon-baryon interactions, which can be directly applied in nuclear many body studies, such as the
Dirac-Brueckner-Hartree-Fock theory.

In the present work, we report on the recent efforts in constructing relativistic baryon-baryon interactions in chiral perturbation theory, focusing on the nucleon-nucleon interaction with $S=0$
and hyperon-nucleon interaction with $S=-1$. More details can be found in Refs.~\cite{Ren:2016jna,Li:2016mln}.
\section{Theoretical Framework}
\subsection{Covariant power counting}
In the covariant power counting, the small scale $Q$ can be either the pseudoscalar meson masses or
the three momentum of the external baryon, and the hard scale is $\Lambda_\chi \sim 1$ GeV.
Employing the naive dimensional analysis, one can determine the chiral dimension ($n_\chi$) of a Feynman diagram,
\begin{equation}
   n_\chi = 4 L - 2N_\phi - N_B +\sum\limits_k k V_k,
 \end{equation}
 where $N_\phi$ ($N_B$) denotes the number of internal meson (baryon) propagators, $L$ the number of loops, and $V_k$ is the number of
 vertices from $k$th order Lagrangians. It is worthy noting that the covariant power counting is well defined
 in the meson-meson and meson-baryon sectors. Although several studies have attempted to extend such a power counting to the baryon-baryon
 sector,  it is still not yet systematically formulated up to higher orders. Here, we follow the strategies of Ref.~\cite{Polinder:2006zh}.

\begin{figure}
  \centering
  \includegraphics[width=0.12\textwidth]{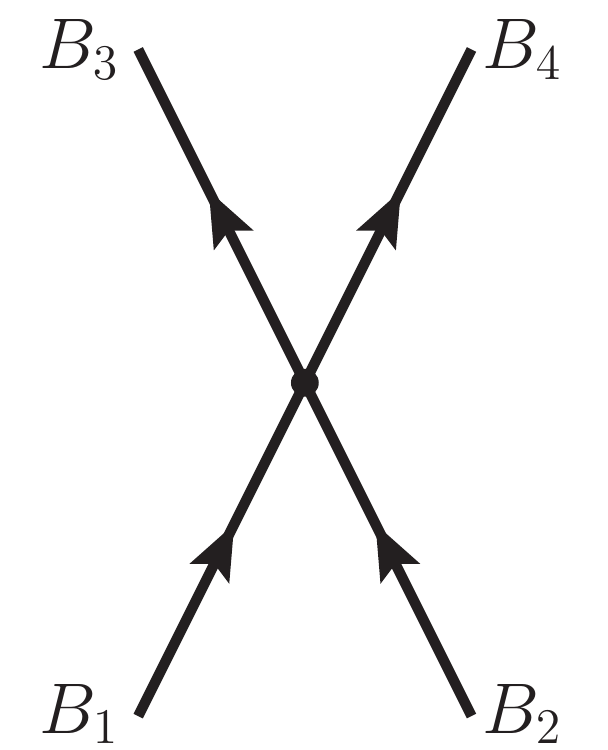} \qquad \qquad
  \includegraphics[width=0.12\textwidth]{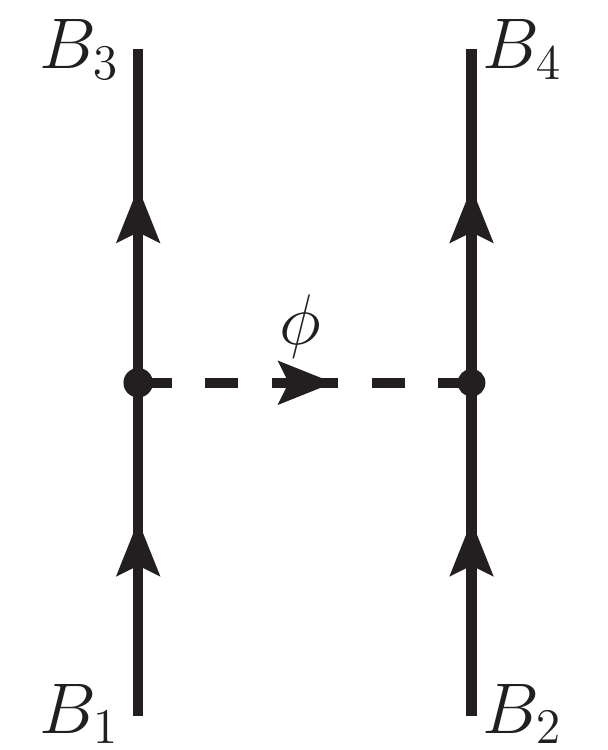}
  \caption{Nonderivative four-baryon contact terms and the OME potential at LO.
    The solid lines denote incoming and outgoing baryons ($B_{1,2,3,4}$), and the dashed line
    represents the exchanged pseudoscalar meson $\phi$.}
  \label{Fig:LO}
\end{figure}

In the current work, we focus on the leading order studies of baryon-baryon interactions, which include
the four-baryon
 contact terms (CT) and one-meson-exchange (OME) contributions, as shown in Fig.~\ref{Fig:LO}.
 The relevant chiral Lagrangians have the following form,
 \begin{equation}
   \mathcal{L}_\mathrm{eff.} =  \mathcal{L}_{\Phi\Phi}^{(2)} + \mathcal{L}_{\Phi B}^{(1)}  + \mathcal{L}_{BB}^{(0)},
 \end{equation}
 with
 \begin{eqnarray}
   &&\mathcal{L}_{\Phi\Phi}^{(2)} = \frac{F_\Phi^2}{4} \langle D_\mu U (D^\mu U)^\dag\rangle
   +\frac{F_\Phi}{4}\langle \chi U^\dag + U \chi^\dag\rangle,\\
   && \mathcal{L}_{\Phi B}^{(1)} =
    \langle\bar B (i\gamma_\mu D^\mu - M_B ) B\rangle
      + \frac{D/F}{2} \langle \bar B \gamma^\mu\gamma_5[u_\mu,B]_\pm \rangle,\\
  &&  \mathcal{L}_{B B}^{(0)} =  \frac{\tilde C_i^1}{2}~
   \left\langle \left(\bar B_a \bar B_b (\Gamma_i B)_b (\Gamma_i B)_a\right)\right\rangle
  + \frac{\tilde C_i^2}{2}~\left\langle\left(\bar B_a (\Gamma_i B)_a \bar B_b (\Gamma_i
     B)_b\right)\right\rangle\nonumber\\
   &&\qquad\quad + \frac{\tilde C_i^3}{2}~\left\langle\left(\bar B_a (\Gamma_i B)_a\right)\textrm{tr}\left( \bar B_b (\Gamma_i B)_b\right)\right\rangle,
 \end{eqnarray}
 where the definition of symbols can be found in Refs.~\cite{Ren:2016jna,Li:2016mln}. Here, we would like to mention that $\Gamma_i$
 ($1,~\gamma^\mu,~\sigma^{\mu\nu}\,~\gamma^{\mu}\gamma_5,~\gamma_5$) are the elements of the Clifford algebra.
 $\tilde{C}_i^m~(m=1,2,3)$ are the
 low-energy constants (LECs) corresponding to independent four-baryon operators.

\subsection{Leading order baryon-baryon interactions}
With the chiral Lagrangians given in the previous subsection, one can easily obtain the chiral baryon-baryon interaction corresponding to the Feynman diagrams of Fig.~\ref{Fig:LO},
 \begin{equation}
    V_\mathrm{LO} = V_\mathrm{CT} + V_\mathrm{OME},
 \end{equation}
 which, as usual, can be divided into four sectors with different strangeness number: $S=0,-1,-2,-3,-4$.
 In the nucleon-nucleon sector $S=0$, only five LECs contribute, while 12 contribute to the $S=-1$ sector.

In the present work, we focus on the nucleon-nucleon interaction and hyperon-nucleon interaction.
We note that  the relativistic contact terms
contribute to all the $J\leq 1$ partial
waves, while the relativistic corrections to the OME potential are largely suppressed. More details can be found in Refs.~\cite{Ren:2016jna,Li:2016mln}.

\subsection{Scattering equation}
Now, we are in a position to apply the relativistic baryon-baryon interactions
to describe scattering observables.
Since the  baryon-baryon interactions are non-perturbative at the low-energy region,
one has to insert them into
the Bethe-Salpeter equation (BSE) and iterate them. In practice, it is difficult to solve the Bethe-Salpeter equation exactly and the
three-dimensional reductions of BSE are often used, such as the Blankenbecler-Sugar equation, the Thompson equation,
and the Kadyshevsky equation.
In the present work, we employ the (generalized) Kadyshevsky equation, which reads as
\begin{align}\label{SEK}
  & T_{\rho\rho'}^{\nu\nu',J}(\mbox{\boldmath $p$}',\mbox{\boldmath $p$};\sqrt{s})
  =
   V_{\rho\rho'}^{\nu\nu',J}(\mbox{\boldmath $p$}',\mbox{\boldmath $p$})
   +
  \sum_{\rho'',\nu''}\int_0^\infty \frac{dp''p''^2}{(2\pi)^3} \frac{M_{B_{1,\nu''}}M_{B_{2,\nu''}}~ V_{\rho\rho''}^{\nu\nu'',J}(\mbox{\boldmath $p$}',\mbox{\boldmath $p$}'')~
   T_{\rho''\rho'}^{\nu''\nu',J}(\mbox{\boldmath $p$}'',\mbox{\boldmath $p$};\sqrt{s})}{E_{1,\nu''}E_{2,\nu''}
  \left(\sqrt{s}-E_{1,\nu''}-E_{2,\nu''}+i\epsilon\right)},
\end{align}
for the coupled channel case with different intermediate baryon masses in the $|LSJ\rangle$ basis. The original Kadyshevsky equation can be obtained
by taking the two intermediate baryon masses equal.
It should be noted  that, in such a equation, the internal energy exchange is not zero, therefore, especially for the
OME potential,  retardation effects should be retained.

Furthermore, the chiral potential has to be regularized in order to avoid ultraviolet divergence.
In the present work, we use a separable exponential form factor,
\begin{equation}\label{EF}
  f_{\Lambda_F}(\mbox{\boldmath $p$},\mbox{\boldmath $p$}') = \exp \left[-\left(\frac{\mbox{\boldmath $p$}}{\Lambda_F}\right)^{2n}-\left(\frac{\mbox{\boldmath $p$}'}{\Lambda_F}\right)^{2n}\right] \, ,
\end{equation}
where $n=2$. Such a form factor is not covariant, as a result, the potential is not fully covariant as well.

\section{Results and discussion}
\subsection{Nucleon-nucleon interaction}
\begin{figure}[t]
\centering
\includegraphics[width=0.45\textwidth,keepaspectratio,angle=0,clip]{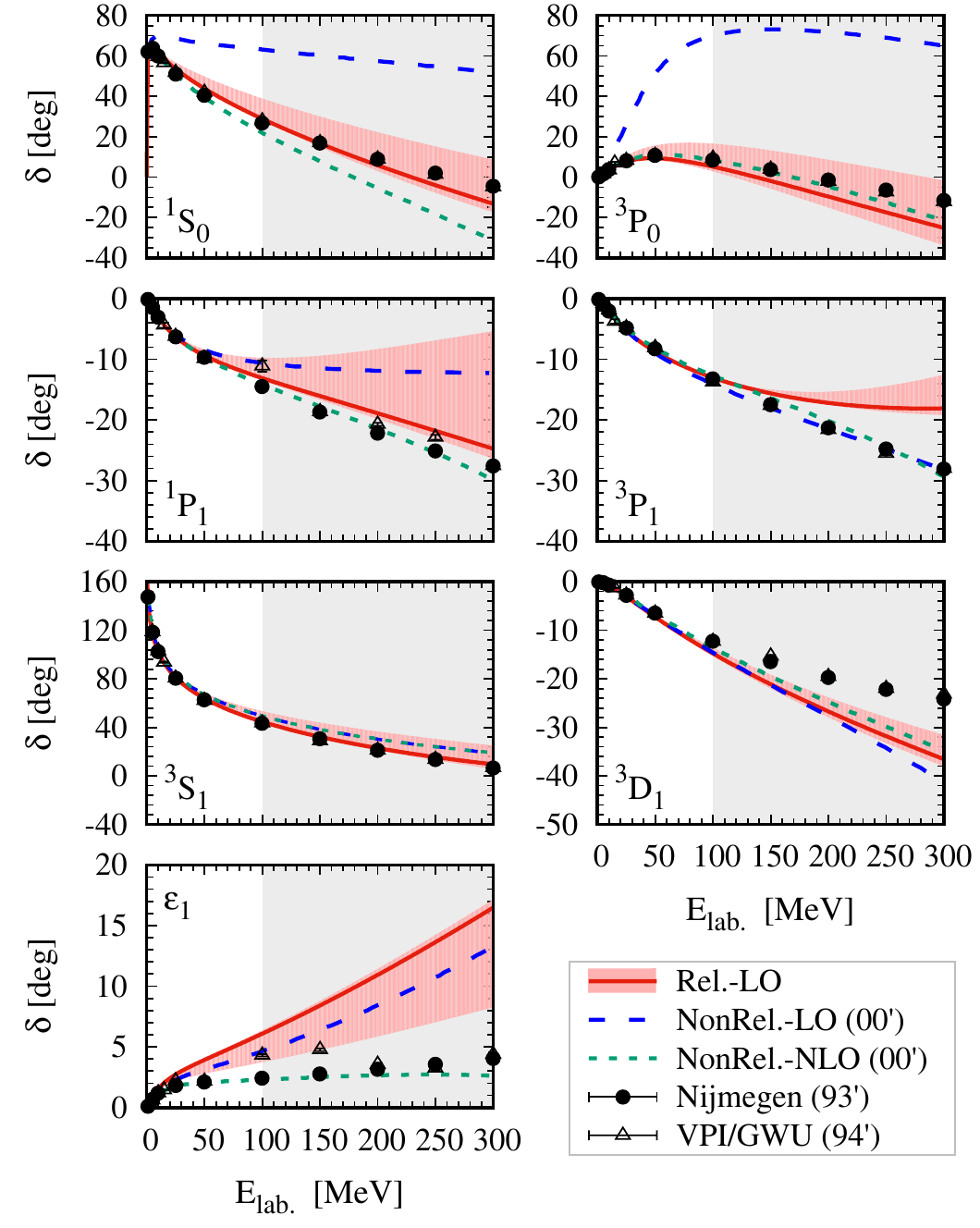}
\caption{Neutron-proton phase shifts for $J\leq1$. The red solid lines denote the relativistic results, while the dashed and dotted lines represent the LO and NLO non-relativistic results~\cite{Epelbaum:1999dj}. The red bands are the relativistic results with the cutoff ranging from 500 MeV to 1000 MeV. Solid dots and open triangles represent the $np$ phase shifts of Nijmegen~\cite{Stoks:1993tb} and VPI/GWU~\cite{Arndt:1994br}. The gray backgrounds denote the energy regions where the  theoretical results are predictions.
\label{Fig:PSdes}
}
\end{figure}

As mentioned before, there are five LECs in the chiral nuclear potential up to LO, which can be fixed by fitting to the
$J=0$, $1$ Nijmegen partial wave phase shifts of the $np$ channel with laboratory kinetic energy, $E_\mathrm{lab},$ $\leq
100$ MeV. By varying the momenta cutoff $\Lambda_F$ from $500$ MeV to $1000$ MeV,
we obtained a minimum of $\tilde{\chi}^2/\mathrm{d.o.f.}\sim 2.0$ at
$\Lambda_F=750$ MeV, and the corresponding LECs are of similar mannitude.
In Fig~\ref{Fig:PSdes}, the description of the Nijmegen and the VPI/GWU phase shifts are shown for $E_\mathrm{lab}$ up
to $300$ MeV.

One can see that the relativistic formulation can describe the $^1S_0$ and $^3P_0$ phase shifts with a quality  similar to those of the NLO non-relativistic chiral force. A variation of the cutoff from
500 MeV to 1000 MeV does not change
qualitatively the overall picture. For the angular momenta $J=1$ waves, the LO relativistic and non-relativistic results
are similar to each other, taking into account the variation of the cutoff.

\subsection{Hyperon-nucleon interaction}

For the hyperon-nucleon potential, the 12 LECs can be determined by fitting to the 36 hyperon-nucleon
scattering data with the following three additional constraints $a^{\Lambda p}_{1S0}=-2.44$ fm, $a^{\Lambda p}_{3S1}=-1.32$ fm
and $a^{\Sigma^+ p}_{3S1} > 0$.
The best fitted $\chi^2$ is around $16$ with a cutoff $\Lambda_F=600$ MeV.

\begin{figure}[h!]
  \centering
  \includegraphics[width=1\textwidth]{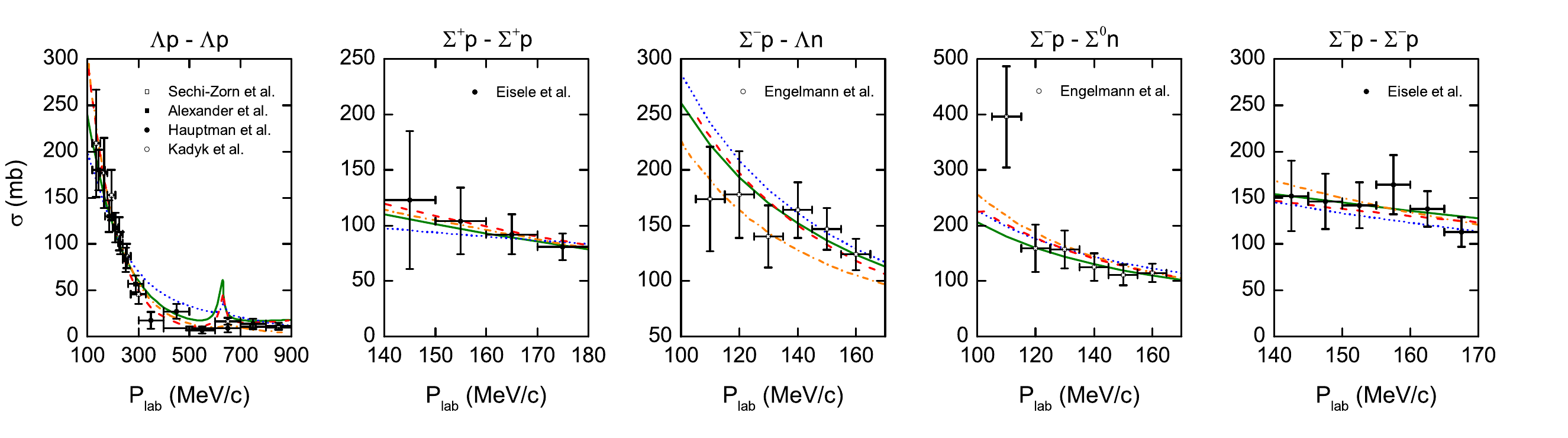}
  \caption{Cross sections in the covariant $\chi$EFT approach (green solid lines) and HB approach (blue dotted lines) at
    LO as functions of the laboratory momentum at $\Lambda_F=600$ MeV. For reference, the NSC97f \cite{Rijken:1998yy}
    (red dash lines) and J\"ulich 04~\cite{Haidenbauer:2005zh} (orange dashed-dotted lines) results are shown.
    The experimental data from different groups are also given. The details can be seen in Ref.~\cite{Li:2016mln}.}
\label{Fig:CrossSection}
\end{figure}

Using the best fitted LECs, in Fig.~\ref{Fig:CrossSection}, we compare the descriptions of the experimental cross
sections with the LO HB approach and those of two phenomenological potentials, NSC97f~\cite{Rijken:1998yy} and J\"ulich
04~\cite{Haidenbauer:2005zh}. One can see that the relativistic potential can reproduce the experimental data rather well. The cusp at the $\Sigma N$ threshold in the $\Lambda p \rightarrow \Lambda p$ reaction is also well reproduced.

\section{Conclusion and outlook}
We have formulated  the relativistic baryon-baryon interactions in covariant chiral perturbation theory up to leading
order. Starting from the Lorentz invariant chiral Lagrangians,
we keep the full form of the Dirac spinors to derive the chiral interactions, which are then iterated using the Kadyshevsky equation.
A rather good description of the $np$ Nijmegen phase shifts and the hyperon-nucleon
scattering data are achieved at leading order, in comparison with the results from the non-relativistic
chiral potentials. It would be interesting to construct the baryon-baryon interactions up to higher orders to test the convergence of covariant chiral perturbation theory in the two-baryon sector.
Furthermore, the relativistic baryon-baryon interactions could also be used as inputs to studies of relativistic (hyper)nuclear structure/reactions, e.g., in the relativistic Dirac-Brueckner-Hartree-Fork theory.

\section*{Acknowledgment}
This work was supported in part by the National Natural Science Foundation of China under Grants No. 11375024, No. 11522539, No. 11335002, and No. 11375120, by DFG and NSFC through funds provided to the Sino-German CRC 110 ¡°Symmetries and the Emergence of Structure in QCD¡± (NSFC Grant No. 11621131001, DFG Grant No. TRR110), the Major State 973 Program of China under Grant No. 2013CB834400, the China Postdoctoral Science Foundation under Grants No. 2016M600845, No. 2017T100008, the Fundamental Research Funds for the Central Universities.


\begin{thebibliography}{99}
\bibitem{Liang:2014dma} 
  H.~Liang, J.~Meng and S.~G.~Zhou,
  Phys.\ Rept.\  {\bf 570}, 1 (2015).
  
\bibitem{Weinberg:1978kz} 
  S.~Weinberg,
  Physica A {\bf 96}, 327 (1979).

  
\bibitem{Weinberg:1990rz}
  S.~Weinberg,
  Phys.\ Lett.\ B {\bf 251}, 288 (1990).


\bibitem{Weinberg:1991um}
  S.~Weinberg,
  Nucl.\ Phys.\ B {\bf 363}, 3 (1991).


\bibitem{Bedaque:2002mn} 
  P.~F.~Bedaque and U.~van Kolck,
  Ann.\ Rev.\ Nucl.\ Part.\ Sci.\  {\bf 52}, 339 (2002).


\bibitem{Epelbaum:2008ga} 
  E.~Epelbaum, H.~W.~Hammer and U.~G.~Mei\ss ner,
  Rev.\ Mod.\ Phys.\  {\bf 81}, 1773 (2009).

\bibitem{Machleidt:2011zz} 
  R.~Machleidt and D.~R.~Entem,
  Phys.\ Rept.\  {\bf 503}, 1 (2011).
  
  
\bibitem{Ren:2016jna}
  X.~-L.~Ren, K.~-W.~Li, L.~S.~Geng, B.~-W.~Long, P.~Ring and J.~Meng,
  arXiv:1611.08475 [nucl-th].
  
\bibitem{Li:2016mln}
  K.-W.~Li, X.-L.~Ren, L.-S.~Geng and B.~Long,
  arXiv:1612.08482 [nucl-th].

\bibitem{Polinder:2006zh} 
  H.~Polinder, J.~Haidenbauer and U.~G.~Mei\ss ner,
  Nucl.\ Phys.\ A {\bf 779}, 244 (2006).
  
\bibitem{Epelbaum:1999dj}
  E.~Epelbaum, W.~Gloeckle and U.~G.~Mei\ss ner,
  Nucl.\ Phys.\ A {\bf 671}, 295 (2000).

\bibitem{Stoks:1993tb}
  V.~G.~J.~Stoks, R.~A.~M.~Klomp, M.~C.~M.~Rentmeester and J.~J.~de Swart,
  Phys.\ Rev.\ C {\bf 48}, 792 (1993).
  
\bibitem{Arndt:1994br}
  R.~A.~Arndt, I.~I.~Strakovsky and R.~L.~Workman,
  Phys.\ Rev.\ C {\bf 50}, 2731 (1994).


\bibitem{Rijken:1998yy}
  T.~A.~Rijken, V.~G.~J.~Stoks and Y.~Yamamoto,
  Phys.\ Rev.\ C {\bf 59}, 21 (1999).
  
\bibitem{Haidenbauer:2005zh}
  J.~Haidenbauer and U.~-G.~Mei\ss ner,
  Phys.\ Rev.\ C {\bf 72}, 044005 (2005).






  
\end{thebibliography}
\end{document}